\documentclass[letterpaper, 10 pt, conference]{ieeeconf}  

\IEEEoverridecommandlockouts                              

\usepackage{graphicx}
\usepackage{booktabs}
\usepackage{amssymb,amsfonts,mathrsfs,amsmath}
\usepackage{multicol}
\usepackage{multirow}
\usepackage{algorithmic}
\usepackage{algorithm}
\usepackage{array}
\usepackage[caption=false,font=normalsize,labelfont=sf,textfont=sf]{subfig}
\usepackage{textcomp}
\usepackage{stfloats}
\usepackage{url}
\usepackage{verbatim}
\usepackage{balance}
\usepackage{cite}
\usepackage{colortbl}
\usepackage{bm}
\usepackage{hyperref}

\newtheorem{remark}{Remark}

\newtheorem{mydef}{Definition}
\newcommand{\bx}{{\bm{x}}}
\newcommand{\by}{{\bm{y}}}
\newcommand{\bM}{{\mathcal{M}}}

\newcommand{\bR}{\mathbb{R}}

\newcommand{\bC}{\mathbb{C}}

\newcommand{\hF}{\mathcal{F}}   
\newcommand{\hM}{\mathcal{M}}

\newcommand{\sspan}{\mathrm{span}}

\title{\LARGE \bf
Koopman Spectral Analysis from Noisy Measurements based on Bayesian Learning and Kalman Smoothing
}

\author{Zhexuan Zeng, Jun Zhou, Yasen Wang, and Zuowei Ping
\thanks{*This work was supported by the National Natural Science Foundation of China under Grant 92167201 and National Scholarship Fund of China.}
\thanks{Zhexuan Zeng is with School of Artificial Intelligence and Automation, Huazhong University of Science and Technology, Wuhan, China.}
\thanks{Jun Zhou is with School of Artificial Intelligence and Automation \& China-EU Institute for Clean And Renewable Energy, Huazhong University of Science and Technology, Wuhan, China}
\thanks{Yasen Wang is with China Telecom Research Institute, Guangzhou, China.}
\thanks{Zuowei Ping is with the with the National Key Laboratory of Electromagnetic Energy, Naval University of Engineering, Wuhan, China.}
\thanks{For correspondence, \href{mailto: pingzuowei@hust.edu.cn}{\tt pingzuowei@hust.edu.cn}.}%
}%

\begin{document}

\maketitle
\thispagestyle{empty}
\pagestyle{empty}

\begin{abstract}
		Koopman spectral analysis plays a crucial role in understanding and modeling nonlinear dynamical systems as it reveals key system behaviors and long-term dynamics. However, the presence of measurement noise poses a significant challenge to accurately extracting spectral properties. In this work, we propose a robust method for identifying the Koopman operator and extracting its spectral characteristics in noisy environments. To address the impact of noise, our approach tackles an identification problem that accounts for both systematic errors from finite-dimensional approximations and measurement noise in the data. By incorporating Bayesian learning and Kalman smoothing, the method simultaneously identifies the Koopman operator and estimates system states, effectively decoupling these two error sources. The method's efficiency and robustness are demonstrated through extensive experiments, showcasing its accuracy across varying noise levels.
\end{abstract}

\section{INTRODUCTION}
The spectral properties of the Koopman operator are fundamental to both the theoretical analysis and data-driven modeling of nonlinear dynamical systems. Specifically, the real and imaginary components of the Koopman spectrum correspond to essential dynamical features -- growth or decay rates and oscillation frequencies, respectively. These spectral characteristics have been shown to be critical for analyzing system stability \cite{mauroy2016global} and determining appropriate frequency bounds for data sampling in nonlinear systems \cite{zeng2024sampling,zeng2024generalized}. Moreover, Koopman spectral analysis enables the decomposition of complex fluid dynamics into distinct Koopman modes, providing a powerful framework for modeling and prediction \cite{mezic2013analysis}. In recent years, this methodology has been successfully applied across a variety of fields, including fluid mechanics \cite{rowley2009spectral, mezic2013analysis}, robotics \cite{bruder2020data}, and power systems \cite{susuki2013nonlinear}.

The widespread adoption of Koopman operator-based methods, first defined in 1931 \cite{Koopman1931hamiltonian}, has been largely driven by the development of the numerical approaches named dynamic mode decomposition (DMD) \cite{schmid2010dynamic}, which enables the identification of the Koopman operator from data. This advancement has spurred extensive research into accurately approximating the Koopman operator, as introduced in the survey \cite{schmid2022dynamic}. To capture the complex dynamical behavior of nonlinear systems more effectively, the extended DMD (EDMD) method \cite{williams2015data} leverages nonlinear observable functions, while neural network-based methods apply deep learning to discover optimal observable functions \cite{yeung2019learning}. However, extracting the spectral properties of the Koopman operator in a noisy environment remains a significant challenge due to its sensitivity to measurement noise \cite{dawson2016characterizing}. To address this, recent efforts have focused on improving Koopman operator approximations in noisy environments. Techniques like forward-backward DMD (fbDMD) \cite{dawson2016characterizing}, which combines results from both forward and backward DMD, and Total Least Squares DMD (TDMD), which utilizes an augmented snapshot matrix to account for noise in all measurements \cite{hemati2017biasing}, have shown promise. Additionally, state estimation methods such as the Kalman filter and extended Kalman filter have been integrated into DMD to mitigate noise effects \cite{nonomura2018dynamic, jiang2022correcting, nonomura2019extended}. While these approaches can effectively correct for noise in certain scenarios, many are constrained by parameter selection, which often depends on prior knowledge of the noise characteristics. Therefore, it is essential to capture and then utilize the statistical properties of measurement noise to accurately extract the Koopman operator’s spectral properties.

In this work, we present a novel algorithm for extracting the spectral properties of the Koopman operator through the iterative refinement of both the identification of the Koopman operator and the estimation of noisy data. By employing time-delay observable functions, we reformulate the problem as the identification of a linear system subject to both systematic and measurement errors. This identification process utilizes Bayesian learning and Kalman smoothing techniques. Specifically, the integration of Bayesian learning facilitates a probabilistic framework for addressing uncertainties inherent in the data, while Kalman smoothing offers an optimal recursive solution for state estimation given the model and noise distribution. Hence, by combining them together, it enables the iterative updating of noise distributions and state estimations, thereby enhancing the accuracy of identifying both the Koopman operator and the system states in noisy environments. It will be demonstrated that, this approach has significant improvements in Koopman spectral analysis compared to some popular noisy dynamic mode decomposition methods, particularly in scenarios where prior knowledge regarding noise levels is limited. 

The rest of this paper is organized as follows. In Section \ref{sec:PF}, we introduce the Koopman operator theory briefly and formulate the noisy spectral analysis problem. Next, we propose the Koopman-Bayesian-Kalman Smoothing method (KBK) to approximate Koopman spectrum in Section \ref{sec:method}. Finally, in Section \ref{sec:experiments}, we demonstrate the effectiveness of the proposed approach through its application to specific nonlinear systems. 

	\subsection{Notation}
	Throughout this paper, we adopt the following notations: The domain of the operator is denoted as $\mathcal D(\cdot)$. The linear space spanned by basis functions is denoted as $\text{span}\{\cdot\}$. The transpose of $a$ is denoted as $a^T$. The Gaussian distribution is denoted as $\mathcal N(a,b)$ with the mean $a$ and covariance $b$. The expectation is denoted as $\mathbb E(\cdot)$. The trace of a matrix $A$ is denoted as $\text{Tr}(A)$. Finally, the probability density function of random variable $x$, the conditional distribution of $x$ given $y$ are denoted as $p(x)$ and $p(x|y)$, respectively. 
	
	\section{Preliminary and Problem formulation}\label{sec:PF}
	\subsection{Koopman operator theory}
	Let us consider an autonomous nonlinear dynamical system:
	\begin{equation}\label{eq1}
		\begin{aligned}
			\dot{\bx}&=f (\bx),
		\end{aligned}
	\end{equation}
	where $\bx\in \bM\subseteq \bR^n$ represents the state vector and $ \bM$ represents the state space. The flow induced by this system is denoted as $S^t,t>0$, i.e., $\bx(t)=S^t(\bx(0))$. The Koopman operator $U^t:\mathcal F\to\mathcal F$ is a linear operator acting on the observable functions of the states, i.e., $g\in\hF:\hM\to\bC$, which is defined as:
    \begin{equation}\label{defkoopman}
		U^t g = g\circ S^t,
	\end{equation} where $\circ$ denotes the composition of two functions.
	
	The infinitesimal generator $L$ of the Koopman operator is defined as \begin{equation}
		Lg = \lim_{t\to 0^+}\frac{1}{t}(U^t-I)g, ~g\in\mathcal D(L),
	\end{equation} where $\mathcal D(L)$ denotes the domain of $L$. The generator is also a linear operator. Assuming that observable functions $g\in\mathcal F$ are continuously differentiable with compact support, we have \begin{equation}
		 L = f\cdot \nabla
	\end{equation} 
	When the generator is bounded, there is an exponential relationship between the Koopman operator and its generator, i.e., \begin{equation}\label{exp}
		U^t = \exp{(Lt)}.
	\end{equation}

     The eigenvalue and associated eigenfunction of the Koopman operator are defined as follows.
	\begin{mydef}[Koopman eigenvalue and eigenfunctions]
		The value $\lambda_i\in\mathbb C$ is an eigenvalue of the Koopman operator if 
		\begin{equation}
			U^t\phi_i = e^{\lambda_it}\phi_i,
		\end{equation}
		where $\phi_i\in\mathcal F,\phi_i\neq0$ is the associated Koopman eigenfunction.
	\end{mydef}
    Noted that the eigenvalue of the Koopman operator $\lambda_i$ is an eigenvalue of the generator with the associated eigenfunction $\phi_i$, i.e., $L \phi_i = \lambda_i\phi_i$. The real and imaginary parts of the eigenvalues describe the growth/decay rate and the oscillation frequency of the observable functions' evolution, which are strongly related to dynamical properties of \eqref{eq1}.
	
 Due to its linearity and rich theoretical support, the Koopman operator theory enjoys wide popularity in nonlinear system identification and control. However, in spite of its success, it is still challenging to consider noise in Koopman operator framework because the statistical property became elusive after lifting from state space to functional space spanned by nonlinear functions.
 
    \subsection{Problem formulation}
	Consider a nonlinear dynamical system described as \eqref{eq1}. The measurements $\{\tilde{\bx}(kT_s)\}_{k=0}^{N-1}$ are uniformly sampled with the sampling period $T_s>0$ and the data size $N$. Then we define $\bm y = [\tilde{\bx}(0),\ldots,\tilde{\bx}((N-1)T_s)]^T$ and $\bm x = [\bx(0),\ldots,\bx((N-1)T_s)]^T$, where $\by$ and $\bx$ respectively represent the noisy measurements and noise-free states. In this paper, we consider an additive zero-mean Gaussian measurement noise $\bm w\in\bR^{N}$, i.e., 
	\begin{equation}
		\bm y = \bm x+ \bm w.
	\end{equation} 
While the Koopman spectral properties can be accurately extracted using DMD methods across a broad range of systems, the presence of measurement noise typically obscures accurate spectral analysis. This highlights the critical need for the development of a robust methodology capable of decoupling the noise component, $\bm w$, from the measurements and ultimately providing a reliable approximation of the Koopman spectrum.

 In this paper, we propose a new approach to extract the Koopman spectral properties of \eqref{eq1} and estimate the states $\{\bx(kT_s)\}_{k=0}^{N-1}$ from its noisy measurements simultaneously.

	\section{Description of proposed method: KBK approach}\label{sec:method}
    The proposed method integrates Kalman smoothing and Bayesian learning within the Koopman operator framework to statistically decouple noise from measurements and approximate the corresponding Koopman spectrum. Specifically, by selecting an appropriate functional space, the spectral analysis problem is reformulated as a linear system identification and state estimation problem. To address these two challenges simultaneously, the Expectation-Maximization (EM) algorithm is employed, combining the strengths of Kalman smoothing for state estimation and Bayesian learning for system identification. Ultimately, the spectrum of the Koopman operator is approximated via the exponential relationship between the Koopman operator and its generator.
	
    The proposed method is outlined in three steps, with the second step inspired by the identification algorithm described in \cite{wang2023data}. A key distinction from \cite{wang2023data} is the absence of a sparsity prior in this work, as the finite-dimensional matrix of the Koopman operator is generally not sparse. 

	\subsection{Koopman operator-based reconstruction framework}
	Theoretically, it is necessary to consider a Koopman-invariant space such that the Koopman operator is well-defined, i.e., $\forall g\in\mathcal F, \ U^\tau g\in\mathcal F,$ which is typically infinite-dimensional. To apply the Koopman operator theory in a numerical framework, we lift the original system to a finite-dimensional functional subspace $\mathcal F_M\subseteq\mathcal F$ spanned by $M$ basis functions. In this work, we choose time-delay functions as basis functions, i.e., $\mathcal F_M=\text{span} \{g(\bx),\ldots,U^{(M-1)T_s}g(\bx)\}$, where $g(\bx) = x_k,$ where $k=1,\ldots,n$, for the following two considerations.
	\begin{itemize}
		\item Since there is no prior knowledge regarding the basis functions $g(\bx)$, selecting a set of analytic functions that ensures the Koopman invariance of $\mathcal F_M$ is challenging. Based on the definition of the Koopman operator $U^t$, the space spanned by time-delay functions is approximately invariant under $U^t$ \cite{brunton2017chaos}.
		\item When the space $\mathcal F_M$ is spanned by time-delay functions, the statistical property of $g\in\mathcal F_M$ are represented clearly because the time-shift transformation does not alter the original distribution of measurement noise, which facilitates statistical analysis.
	\end{itemize}
	In the following, we utilize the observable function $g(\bx) = x_1$ as an example, though this choice is not unique. For instance, we could alternatively select both $g_1(\bx) = x_1, g_2(\bx)=x_2$. Then we arrange the noise-free and noisy samples of $g(t)$ into the following matrices $X,Y\in \bR^{Q\times M}$, respectively: \begin{equation}
		X = \left(\begin{matrix}
			x_1(0)& \ldots & x_1((M-1)T_s)\\
			\vdots & \ddots &\vdots\\
			x_1((Q-1)MT_s) &\ldots&x_1((QM-1)T_s)
		\end{matrix}
		\right),
	\end{equation}
	\begin{equation}
		Y = \left(\begin{matrix}
			\tilde{x}_1(0)& \ldots & \tilde{x}_1((M-1)T_s)\\
			\vdots & \ddots &\vdots\\
			\tilde{x}_1((Q-1)MT_s) &\ldots&\tilde{x}((QM-1)T_s)
		\end{matrix}
		\right).
	\end{equation} We denote the transpose of the $k$-th row of $X,Y$ as $\bm z_k$ and $\bm y_k$ respectively, i.e., for $k=1,\ldots,Q$, \begin{align}
		&\bm z_k = [x_1((k-1)MT_s),\ldots,x_1((kM-1)T_s)]^T,\\
		&\bm y_k = [\tilde{x}_1((k-1)MT_s),\ldots,\tilde{x}_1((kM-1)T_s)]^T.
	\end{align} Denote $A\in\mathbb R^{M\times M}$ as the transpose of the matrix representation of $U^{MT_s}$, i.e., $U^{MT_s}[g(t),\ldots, g(t+(M-1)T_s)] = [g(t),\ldots, g(t+(M-1)T_s)]A^{T}$. Based on the definition of the Koopman operator, we have \begin{equation}\label{reproblem}
	    \begin{aligned}
		\bm z_{k+1} &= A \bm z_k+ \bm v_k,\\
		\bm y_k &= \bm z_k+ \bm w_k,
	\end{aligned}
	\end{equation} where $\bm v_k$ models the error caused by the projection of $U^{MT_s}g$ to $\mathcal F_M$, $\bm w_k$ denotes the measurement noise of samples. We assume that  $\bm w_k\sim \mathcal N(0,R_w), \bm v_k\sim\mathcal N(0,R_v),$ where $R_w,R_v$ are diagonal matrices. 
 \begin{remark}
    The finite-dimensional projection error $\bm v_k$ is modeled as a Gaussian distribution random error in this study, which is motivated by several considerations. First, it has been demonstrated that in complex nonlinear systems, while the projection error may inherently be deterministic, it often exhibits random-like behavior when the system is represented in finite-dimensional linear ways \cite{brunton2017chaos}. Furthermore, each component of $\bm v_k$ tends to remain close to zero (small positive or negative values) due to the approximate invariance of the chosen time-shift space $\hF_M$. From a numerical perspective, modeling this projection error as Gaussian-distributed random noise simplifies the analysis and allows for the effective application of statistical inference tools, thereby streamlining the modeling and analytical process. 
 \end{remark}
	
	Correspondingly, the spectral analysis problem with noisy measurements is transformed to a linear system identification and state estimation task. Specifically, the goal is to identify the matrix and estimate states $\{\bm z_k\}_{k=1}^Q$ from measurements $\{\bm y_k\}_{k=1}^Q$.
	
	
	
	\subsection{Matrix identification and sample estimation}\label{sec:iden_est}
	
	Here we employ the expectation–maximization (EM) framework to solve the identification problem of the matrix $A$ and the estimation of $\bm x_k$, which combines the technique of Bayesian learning and the Kalman smoothing. 
	
	We aim to maximize the likelihood function of observed samples, i.e., \begin{equation}
		\begin{aligned}
			&p(Y|X;A,\bm m_{1:Q},P_{1:Q}, R_v,R_w) \\= &\int p(Y,X;A,\bm m_{1:Q},P_{1:Q},R_v,R_w){\rm d} X,
		\end{aligned}
	\end{equation} where $\bm m_k$ and $P_k$ denotes the mean and covariance of $\bm x_k$, respectively. Let us denote $\bm \theta = \{A,\bm m_{1:Q},P_{1:Q},R_v,R_w\}$. In the framework of EM algorithm, we maximize the following objective function to optimize the parameters $\bm \theta$: \begin{equation}\label{objective}
		\mathbb E_{X|Y;\bm \theta}[\log p(Y,X; \bm \theta)].
	\end{equation}
	We derived the iterative formula of parameters $A,\bm m_{1:Q},P_{1:Q},R_v,R_w$ to solve this problem. Noted that here the matrix $A$ is not assumed to be sparse as \cite{wang2023data} does. To be more specific, we repeat the following steps.
	\begin{itemize}
		\item The $i$-th E-step: 
		We employee Kalman smoothing to estimate $\bm x_k$, where $p(\bm x_k|Y;\bm \theta^i)\sim\mathcal N(\bm m_k^i,P_k^i)$, where $\bm\theta^i$ denotes the $i$th estimation of $\bm\theta$. Initializing the mean and covariance of $\bm x_0$ as $\bm \mu_1^i$ and $\Sigma_1^i$, respectively, we have
		\begin{align}
			\overline{\bm \mu}_k^i &= A^i\bm \mu_{k-1}^i,\\
			\overline{\Sigma}_k^i &= A^i\Sigma_{k-1}^i (A^i)^T+R_v^i,\\
			K_k^i &= \overline{\Sigma}_k^i(\overline{\Sigma}_k^i+R_w^i)^{-1},\\
			\bm \mu_k^i &= \overline{\bm \mu}_k^i+K_k^i(\bm y_k-\overline{\bm \mu}_k^i)\label{mu},\\
			\Sigma_k^i &= (I_M-K_k^i)\overline{\Sigma}_k^i\label{Sigma},
		\end{align} for $k=2,\ldots,Q$.
		Then we use the initial condition $\bm m_Q^i = \bm \mu_Q^i, P_Q^i = \Sigma_Q^i, P_{Q,Q-1}^i = (I_M-K_Q^i)A^i\Sigma_{Q-1}^i$ to compute $\bm m_k^i, P_k^i$ for $k=1,\ldots,Q-1$, and $P_{k,k-1}^i$ for $k=2,\ldots,Q-1$:
		\begin{align}
			&\bm m_k^i = \bm \mu_k^i+G_k^i(\bm m_{k+1}^i-\overline{\bm \mu}^i_{k+1}),\label{mk}\\
			&P_k^i = \Sigma_k^i+G_k^i(P_{k+1}^i-\overline{\Sigma}_{k+1}^i)(G_{k}^{i})^T,\label{pk}\\
			&P_{k,k-1}^i = \Sigma_k^i (G_k^i)^T+G_k^i(P_{k+1,k}-A^i\Sigma_k^i)G_{k-1}^T\label{pkk}
		\end{align} where $G_k^i = \Sigma_k^i (A^i)^T(\overline{\Sigma}_{k+1}^i)^{-1}$.
		
		\item The $i$-th M-step: 
		Based on the mean $\bm m_k^i$ and covariance matrix $P_k^i$ of $\bm x_k$, the maximization problem is reformulated in terms of the following minimization problem: \begin{equation}
			\begin{aligned}
				&\mathop{\arg\min}\limits_{A,R_v,R_w} \mathcal L(A, R_v,R_w) = \sum_{k=1}^{Q} (\log|R_w|+\bm y_k^T R_w^{-1} \bm y_k\\&+\bm m_k^TR_w^{-1}\bm m_k+ \text{Tr}(R_w^{-1}P_k) -2\bm y_k^TR_w^{-1}\bm m_k)\\&+\sum_{k=1}^{Q-1}(\log|R_v|+\bm m_{k+1}^TR_v^{-1}\bm m_{k+1}+\bm m_k^TA^TR_v^{-1} A\bm m_k\\&-2\bm m_{k+1}^T R_v^{-1}A \bm m_k+\text{Tr}(R_v^{-1}P_{k+1}+A^TR_v^{-1}AP_k\\&-2R_v^{-1}AP_{k+1,k})).
			\end{aligned}
		\end{equation}
		Denote $A^{i+1}_j$, $(P_{k+1,k}^i)^j$ and $(\bm m_{k}^i)^j$ as the $j$-th row of $A^{i+1}$, the $j$-th row of $P_{k+1,k}^i$, and the $j$-th component of $\bm m_{k}^i$, respectively. The parameters $A,R_v,R_w$ are optimized by block coordinate decent method, whose iterative formulas are given as follows:\begin{equation}\label{updateRw}
			R_w^{i+1} = \frac{\sum_{k=1}^Q(H_k^i-2\bm m_k^i\bm y_k^T+\bm y_k\bm y_k^T)}{Q},
		\end{equation}
		\begin{equation}\label{updateRv}
			\begin{aligned}
				R_v^{i+1} &= \frac{\sum_{k=1}^{Q-1}\left(A^iH_{k}^i(A^i)^T+H_{k+1}^i\right)}{Q-1}
				\\&-\frac{\sum_{k=1}^{Q-1}2(\bm m_{k+1}^i(\bm m_{k}^i)^T+P_{k+1,k}^i)(A^i)^T}{Q-1},
			\end{aligned}
		\end{equation}
		\begin{equation}\label{updateA}
			\begin{aligned}
				A^{i+1}_j=\left(\sum_{k=1}^{Q-1}((P_{k+1,k}^i)^j+\bm (m_{k+1}^i)^j \bm (m_{k}^i)^T)\right)H_s,
			\end{aligned}
		\end{equation} where $H_s = \left(\sum_{k=1}^{Q-1}H_k^i\right)^{-1}, H_k^i = P_k^i+\bm m_k^i(\bm m_k^i)^T, j=1,\ldots,M$.
	\end{itemize} 
	
	\subsection{Spectral approximation and state estimation}
        Upon convergence of the iterative formulas \eqref{updateRw}-\eqref{updateA}, both the estimated sample values, denoted as $\{\bm m_k\}_{k=1}^Q$, and the approximate matrix representation of $U^{MT_s}$, denoted as $A^T$, are obtained. The spectrum of $U^{MT_s}$ can then be approximated by the eigenvalues of $A^T$. For continuous-time systems, the spectrum of the generator is computed based on the exponential relationship between the generator and the Koopman operator, given by $\log(\lambda)/MT_s$, where $\lambda$ is an eigenvalue of $A$. To enhance the robustness of the approximation, only a subset of eigenvalues of $A$ is selected for the computation of the Koopman spectrum, focusing on those with larger $|\lambda|$. This approach ensures that the selected eigenvalues capture the dominant dynamical features of the system, as modes associated with smaller $|\lambda|$ contribute less significantly to the long-term behavior of the system.

	\begin{remark}[Convergence of optimization]
		The optimization process in KBK method described in Section \ref{sec:iden_est} aims to discover the linear lifted system and estimate the samples from noisy measurements. Let us denote $\bm \theta^* = \{\bm m_{1:Q}^*, P_{1:Q}^*, A^*, R_v^*, R_w^*\}$ as limit point of the sequence $\bm \theta^i 
		$ generated by this algorithm 
  . Based on the theoretical result of \cite{gibson2005robust}, it is guaranteed that the limit point is a stationary point of log-likelihood function $\mathcal P (\bm \theta)$ from any initialization point, where \begin{equation}
			\mathcal P (\bm \theta) = \log p(Y|X;\bm \theta) = \mathcal Q(\bm \theta,\bm \theta') - \mathcal H(\bm \theta,\bm \theta'),
		\end{equation} with $\bm \theta'$ being the current estimate of $\bm \theta$ and \begin{align}
			\mathcal Q(\bm \theta,\bm \theta') = \mathbb E_{X|Y;\bm \theta'}[\log p(Y,X;\bm \theta)],\\
			\mathcal H(\bm \theta,\bm \theta') = \mathbb E_{X|Y;\bm \theta'} [\log p(X|Y;\bm \theta)].
		\end{align}
	\end{remark}

	\section{Numerical experiments}\label{sec:experiments}

In this section, we conduct comprehensive numerical experiments to validate the noise robustness of the proposed Koopman Bilinearization with Kalman Filtering (KBK) method introduced in Section \ref{sec:method}. These experiments are designed to rigorously assess the method's performance in two primary areas: (1) the accurate estimation of Koopman eigenvalues across varying noise levels, and (2) the effective reconstruction of system states under noisy conditions. Additionally, we briefly discuss the experience of selection for the number of data points $N$ and the dimension of the basis $M$ in experiments as a guideline.

To establish a meaningful benchmark, the KBK method is compared with four established approaches: standard Dynamic Mode Decomposition (DMD), total least squares DMD (TDMD) \cite{hemati2017biasing}, forward-backward DMD (fbDMD) \cite{dawson2016characterizing}, and Kalman filter-enhanced DMD (KF-DMD) \cite{jiang2022correcting}. All methods are evaluated on the same datasets with controlled noise levels to ensure a fair comparison.

For quantitative evaluation, two error metrics are employed. First, the relative eigenvalue error is defined as
\begin{equation}
E_{1} = \frac{ \| \bm{\lambda}_{\mathrm{approx}}  -  \bm{\lambda}_{\mathrm{true}} \|_2 }{\| \bm{\lambda}_{\mathrm{true}} \|_2},
\label{eq:error_1}
\end{equation}
where \(\bm{\lambda}_{\mathrm{approx}} \in \mathbb{C}^n\) are the estimated Koopman eigenvalues and \(\bm{\lambda}_{\mathrm{true}} \in \mathbb{C}^n\) represent the theoretically derived eigenvalues. Second, the accuracy of state estimation is measured via the root mean squared error (RMSE):
\begin{equation}
    E_2 =  \sqrt{\frac{1}{N} \sum_{k=0}^{N-1} \left( \hat{x}_1(kT_s) - x_1(kT_s) \right)^2},
    \label{eq:error_2}
\end{equation}
where \(N\) denotes the total number of sampled states, and \(\{\hat{x}_1(kT_s)\}_{k=0}^{N-1}\) and \(\{x_1(kT_s)\}_{k=0}^{N-1}\) are the estimated and true states, respectively.

To provide a comprehensive assessment, three representative nonlinear systems are examined: one with real eigenvalues, one with purely imaginary eigenvalues, and one with complex eigenvalues. These systems are chosen to highlight the versatility and noise robustness of the KBK method in handling a diverse range of dynamical behaviors.

\subsection{Nonlinear system with real Koopman eigenvalues}
We first consider a nonlinear dynamical system described by \eqref{eq:system3}:
\begin{equation}
\begin{cases}
\dot{x}_1 &= -x_1 \\
\dot{x}_2 &= x_1^2 - x_2
\end{cases}
\label{eq:system3}
\end{equation}
This dynamical system can be represented by a finite-dimensional linear model using observable functions ${g_1(\bx) = x_1, g_2(\bx) = x_2, g_3(\bx) = x_1^2}$. Under this representation, the Koopman eigenvalues are $-1$ and $-2$, both of which lie on the real axis. Due to all eigenvalues being negative real numbers, the system's state variables converge monotonically and exponentially to the equilibrium point \((0, 0)\), lacking oscillatory and periodic characteristics. 

\begin{figure*}[!t]
		\centering
		\subfloat[]{
			\label{Fig3.sub1}
			\includegraphics[width=.3\textwidth]{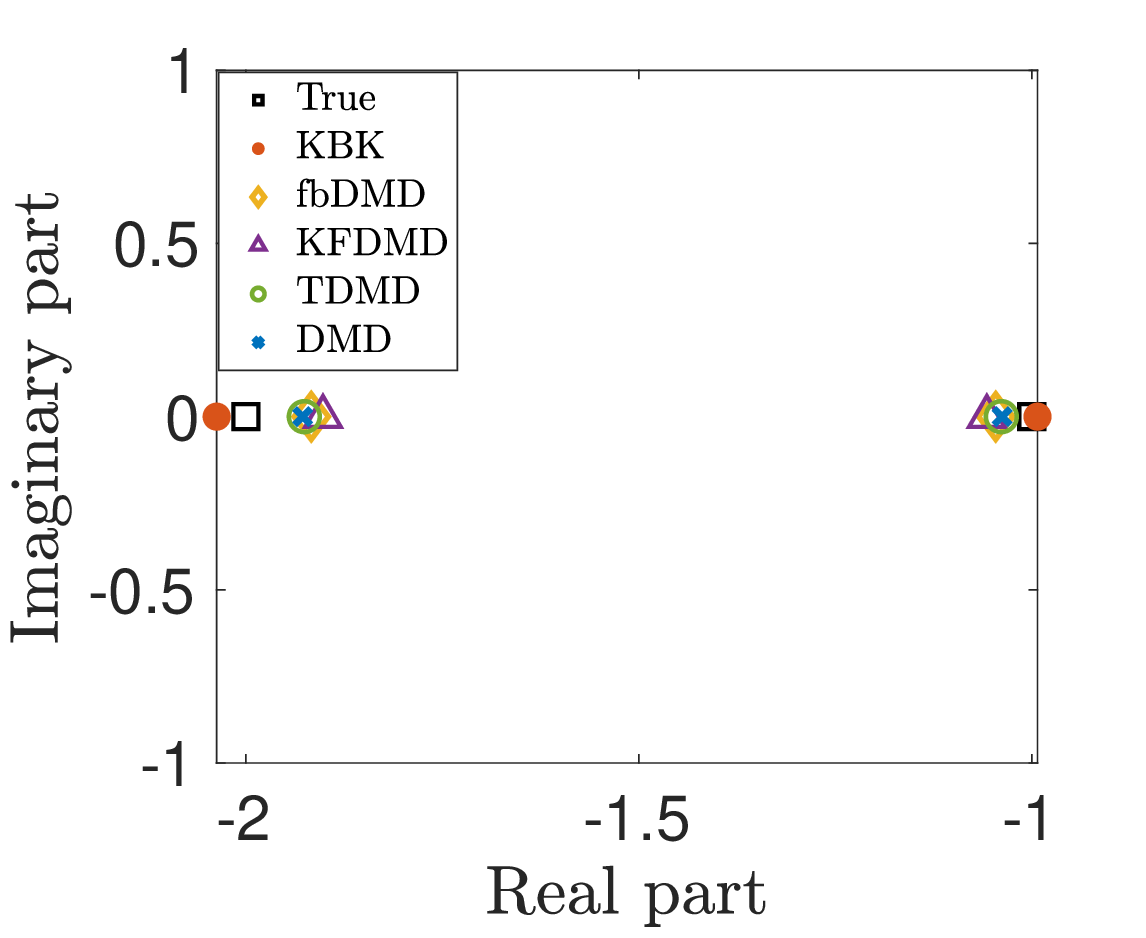}}
		\subfloat[]{
			\label{Fig3.sub2}
			\includegraphics[width=.3\textwidth]{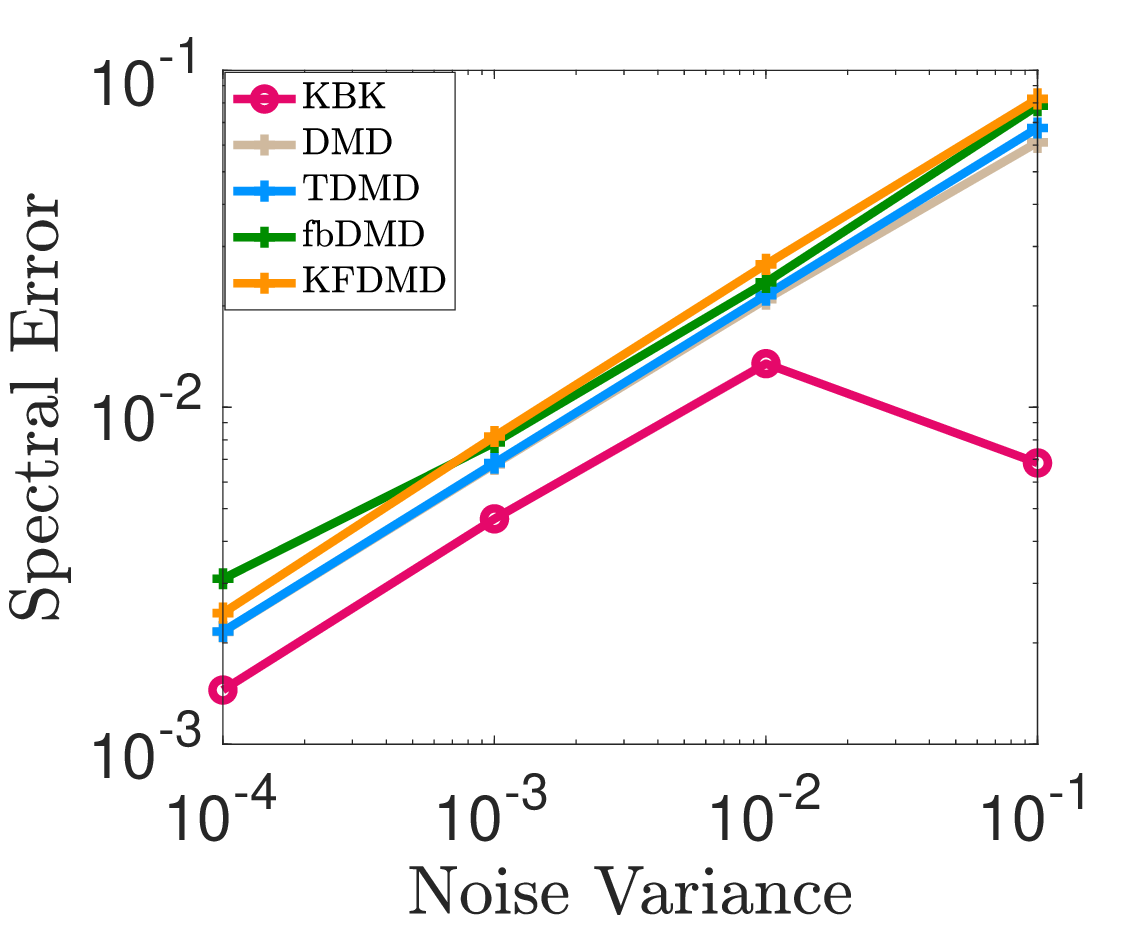}}
		\subfloat[]{
			\label{Fig3.sub3}			\includegraphics[width=.3\textwidth]{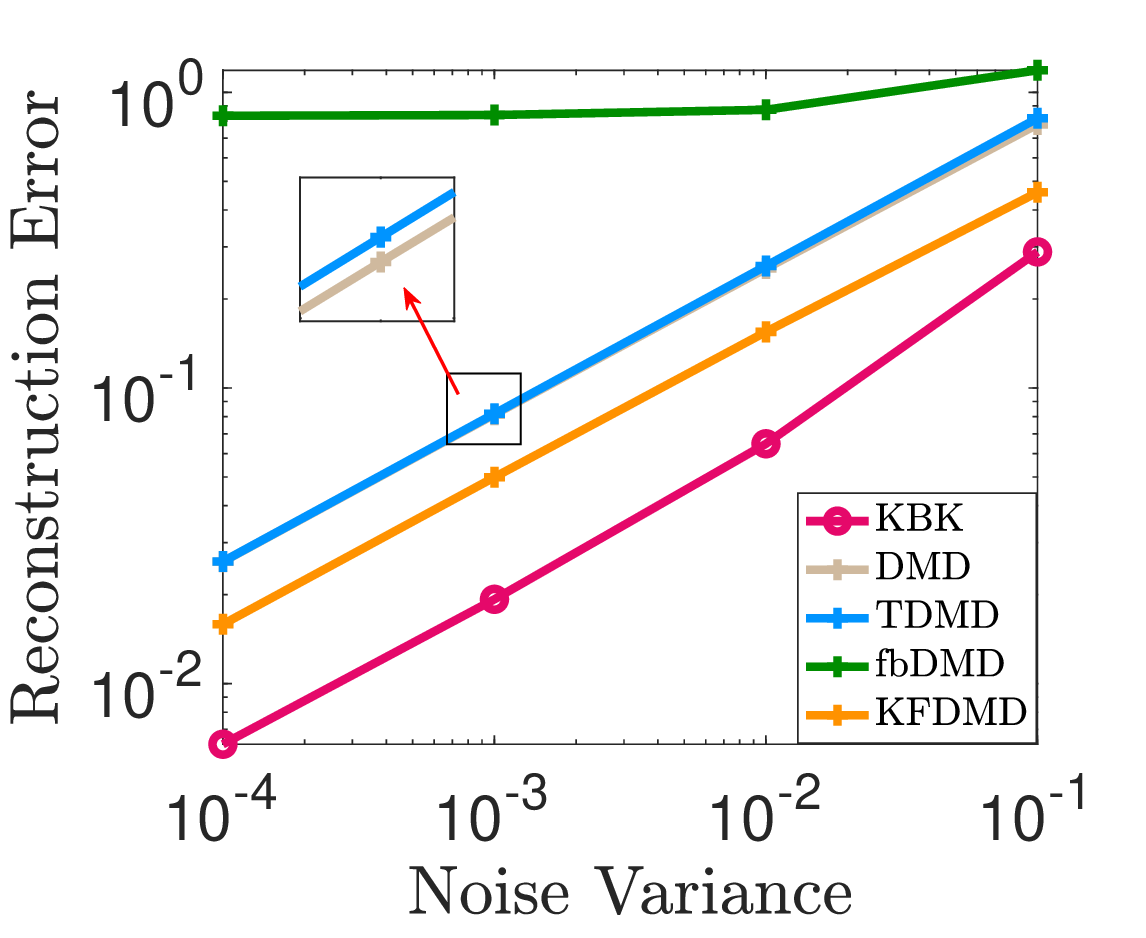}}	
		\caption{The spectral analysis performance of the nonlinear system \eqref{eq:system3} using DMD, TDMD, fbDMD, KFDMD and KBK methods. (a) the spectral approximation when $\sigma^2 = 10^{-2}$, (b) the spectral approximation errors across varying noise variances $\sigma^2$, (c) the state reconstruction errors across varying noise variances $\sigma^2$.}
		\label{fig3}
	\end{figure*}

In this experiment, we utilize $30$ snapshot pairs with additive Gaussian noise, where the noise variance $\sigma^2$ is set to $\{10^{-4}, 10^{-3}, 10^{-2}, 10^{-1}\}$. In the KBK method, we select the observable space $\sspan\{g(\bx), \ldots, U^{3T_s}g(\bx)\}$, where $T_s =0.2$s. The corresponding approximation results are depicted in Fig. \ref{fig3}, demonstrating the impact of varying noise levels on the accuracy of the methods' performance. 

Fig. \ref{Fig3.sub1} illustrates the complex plane of the eigenvalue approximation results using the five methods under a noise variance of $10^{-2}$. The theoretical Koopman eigenvalues of the nonlinear system is indicated by black square markers, while scatter points in various colors represent the approximations computed by different methods. It shows that the imaginary part of the approximated result for all methods are close to zero, aligning well with the true values. However, in terms of the real part, the eigenvalues estimated by the KBK method are significantly closer to the true values, as indicated by the black squares, compared to the other methods. This highlights the superior accuracy of the KBK method in capturing the growth and decay rates of the system's trajectory, making it more reliable for analyzing the dynamical behavior.

Fig.\ref{Fig3.sub2} further reveals the performance differences of various methods under different noise levels. It is important to note that the parameters for each method remain unchanged across the different noise levels in all experiments. As noise variance increases from $10^{-4}$ to $10^{-1}$, the spectrum approximation error (computed using \eqref{eq:error_1}) of the KBK method consistently remains lower than that of other approaches, demonstrating its superior robustness. This advantage stems from the synergistic combination of the Kalman smoothing technique, which enhances the estimation accuracy of data, and the Bayesian framework, which bolsters system identification by accounting for uncertainty in the inference process. As a result, the KBK method maintains accurate spectral analysis even in high-noise environments. We also observe an intriguing phenomenon: at noise variances between $10^{-2}$ and $10^{-1}$, the KBK method exhibits a declining error trend, a pattern that also appears in later experiments with other methods. This may be due to the fact that, at certain noise levels, the methods effectively adjust their balance between accurately estimating states and filtering out noise. In this range, the algorithms may prioritize noise suppression in a way that unexpectedly improves overall performance, resulting in lower estimation errors. In other words, the algorithms appear to exploit higher noise levels to fine-tune their estimation processes, leading to better performance within a specific noise range.

Subsequently, we conducted a comparative analysis of state estimation errors $E_2$ calculated using \eqref{eq:error_2} across various noise levels, as illustrated in Fig.\ref{Fig3.sub3}. The findings reveal that DMD and TDMD exhibit similar error performance in state reconstruction tasks. Although KFDMD also employs the Kalman filter for state estimation, its effectiveness is significantly dependent on precise prior knowledge of the noise level, rendering it less adaptable to varying noise conditions. In contrast, the KBK method consistently demonstrates superior reconstruction performance across all noise levels. This indicates that the KBK method excels not only in spectral analysis but also in state estimation for this system.

 \subsection{Nonlinear system with pure imaginary eigenvalues}
Consider a nonlinear oscillatory system described by $\eqref{eq:system2}$:
\begin{equation}
\begin{cases}
\dot{x}_1 = -x_2 + x_1(1 - x_1^2 - x_2^2), \\
\dot{x}_2= x_1 - x_2(x_1^2 + x_2^2)
\end{cases}
\label{eq:system2}
\end{equation}

For this system, the Koopman eigenvalues are purely imaginary, with the principal eigenvalues being $i$ and $-i$. We utilize $60$ snapshot pairs for numerical experiments with the number of basis $M =4 $ and sampling period $T_s =0.1$s. 

\begin{figure*}[!t]
		\centering
		\subfloat[]{
			\label{Fig2.sub1}
			\includegraphics[width=.3\textwidth]{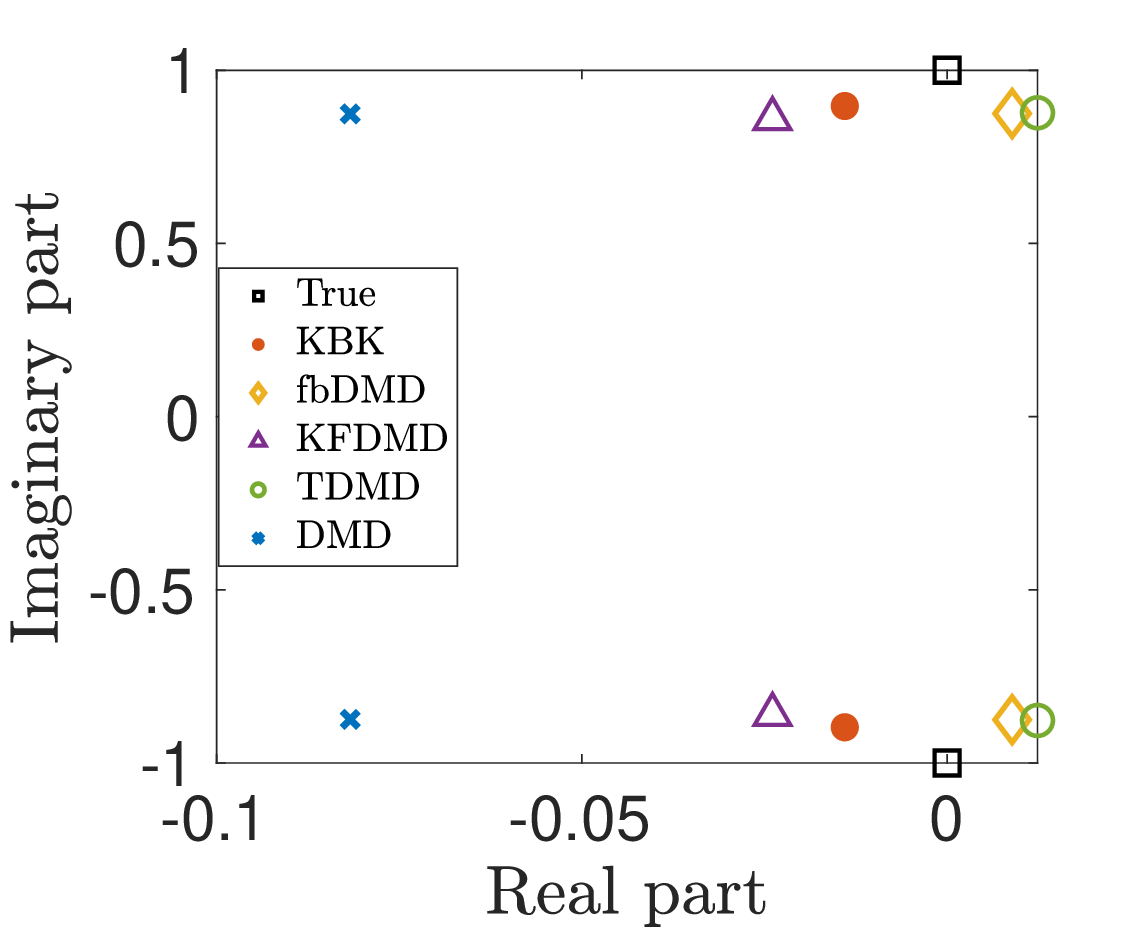}}
		\subfloat[]{
			\label{Fig2.sub2}
			\includegraphics[width=.3\textwidth]{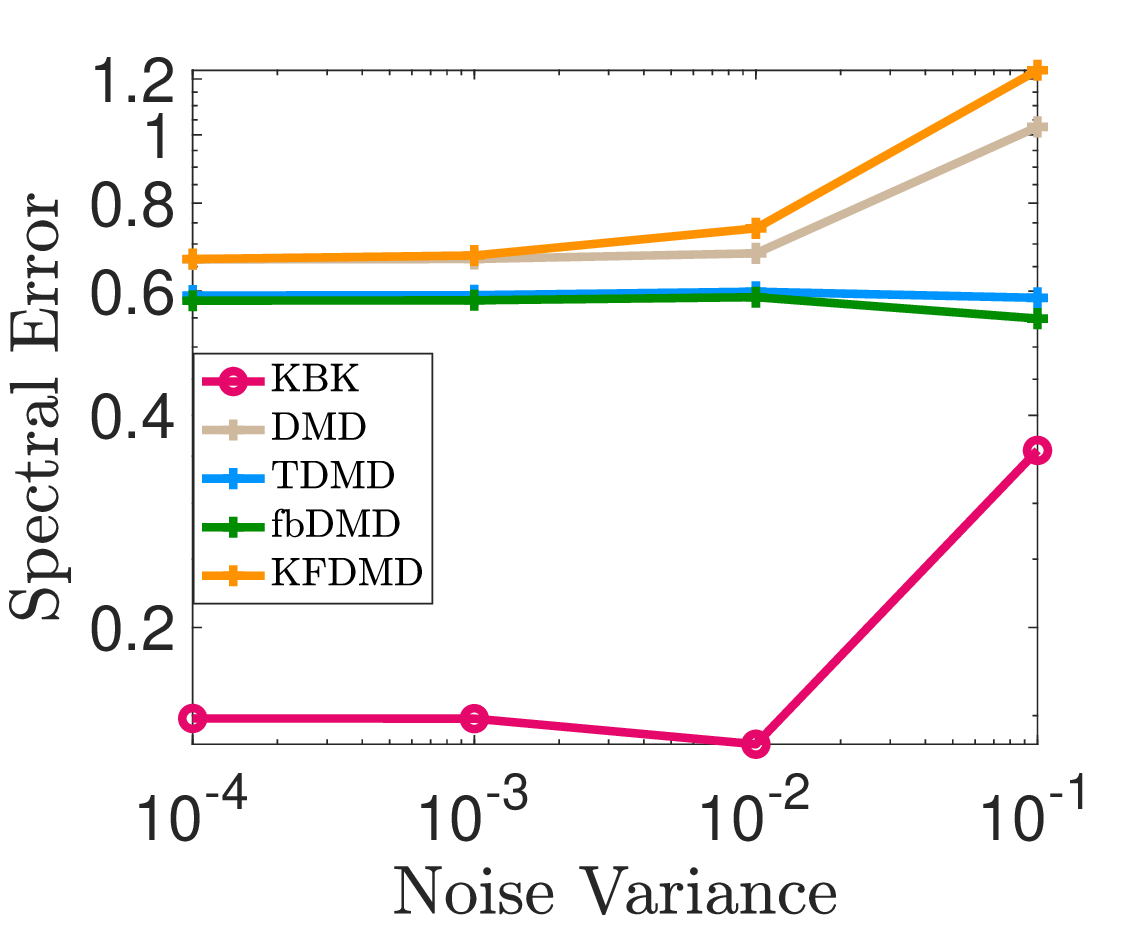}}
		\subfloat[]{
			\label{Fig2.sub3}
			\includegraphics[width=.3\textwidth]{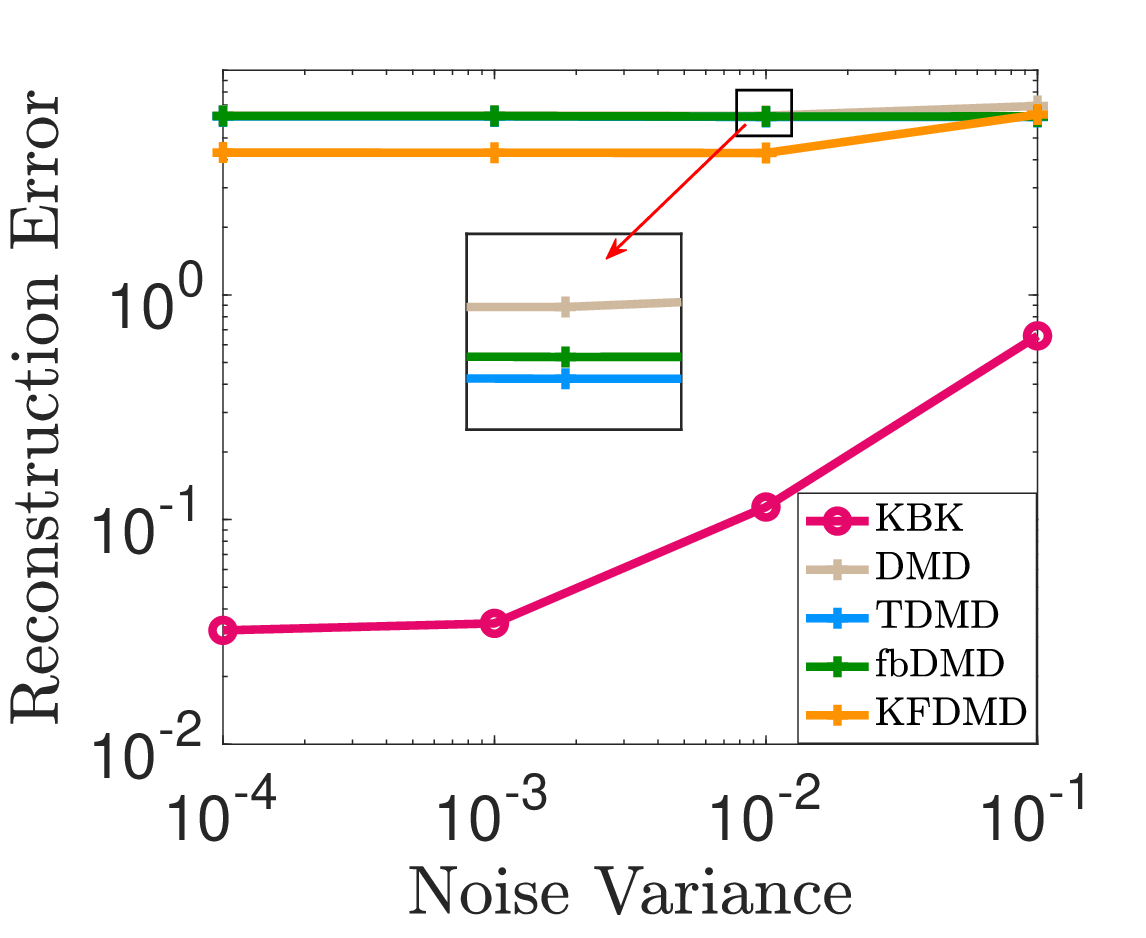}}

		\caption{The spectral analysis performance of the nonlinear system \eqref{eq:system2} using DMD, TDMD, fbDMD, KFDMD and KBK methods. (a) the spectral approximation when $\sigma^2 = 10^{-2}$, (b) the spectral approximation errors across varying noise variances $\sigma^2$, (c) the state reconstruction errors across varying noise variances $\sigma^2$.}
  
		\label{fig2}
	\end{figure*}

Fig. \ref{Fig2.sub1} presents the approximation results under a noise level of $\sigma^2=10^{-2}$. It shows that the proposed KBK method provides the most accurate approximations, closely matching the true eigenvalues and surpassing the other methods in both the real and imaginary components. Furthermore, Fig. \ref{Fig2.sub2} illustrates the approximation error across various noise levels. It shows that the KBK method consistently exhibits significantly lower errors compared to the other methods. 
Additionally, Fig. \ref{Fig2.sub3} presents the state estimation error performance of each method across varying noise variance levels. While the KBK method exhibits greater variability in its effectiveness as noise increases compared to other methods, its error consistently remains lower than that of all other methods under these noise conditions. Interestingly, despite the KFDMD method’s poor performance in spectral analysis for this system, it shows relatively lower error in state estimation compared to DMD, TDMD, and fbDMD methods when the noise level is below $10^{-1}$, indicating its advantage in state estimation. In contrast, the KBK method demonstrates superior accuracy in both spectral analysis and state estimation, as evidenced by this system.

 \subsection{Nonlinear system with complex eigenvalues}

We consider the following nonlinear system whose Koopman eigenvalues admit non-zero real and imaginary parts:

\begin{equation}
\begin{cases}
\dot{x}_1 = -3x_2 - x_1(x_1^2 + x_2^2 + 1), \\
\dot{x}_2 = 3x_1 - x_2(x_1^2 + x_2^2 + 1)
\end{cases}
\label{eq:system1}
\end{equation}

\begin{figure*}[!t]
		\centering
		\subfloat[]{
			\label{Fig1.sub1}
			\includegraphics[width=.3\textwidth]{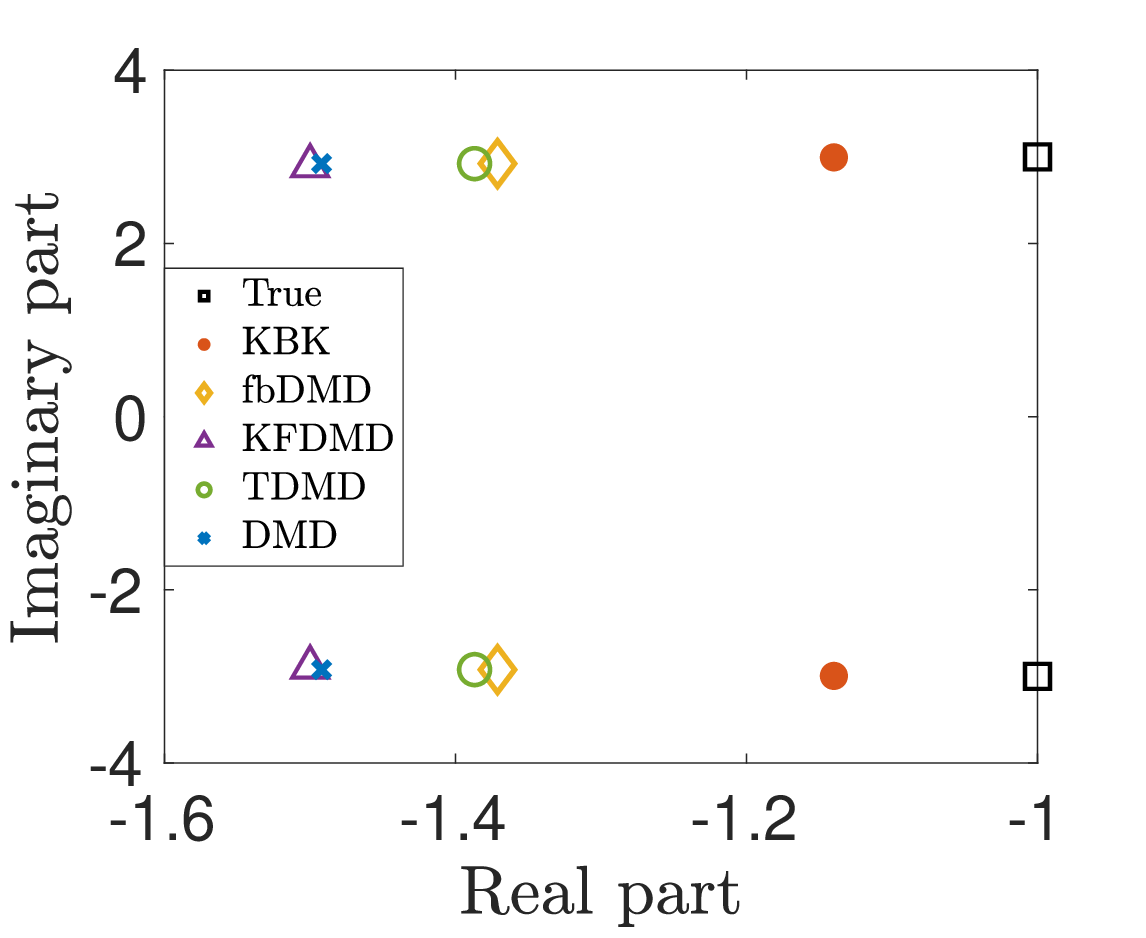}}
		\subfloat[]{
			\label{Fig1.sub2}
			\includegraphics[width=.3\textwidth]{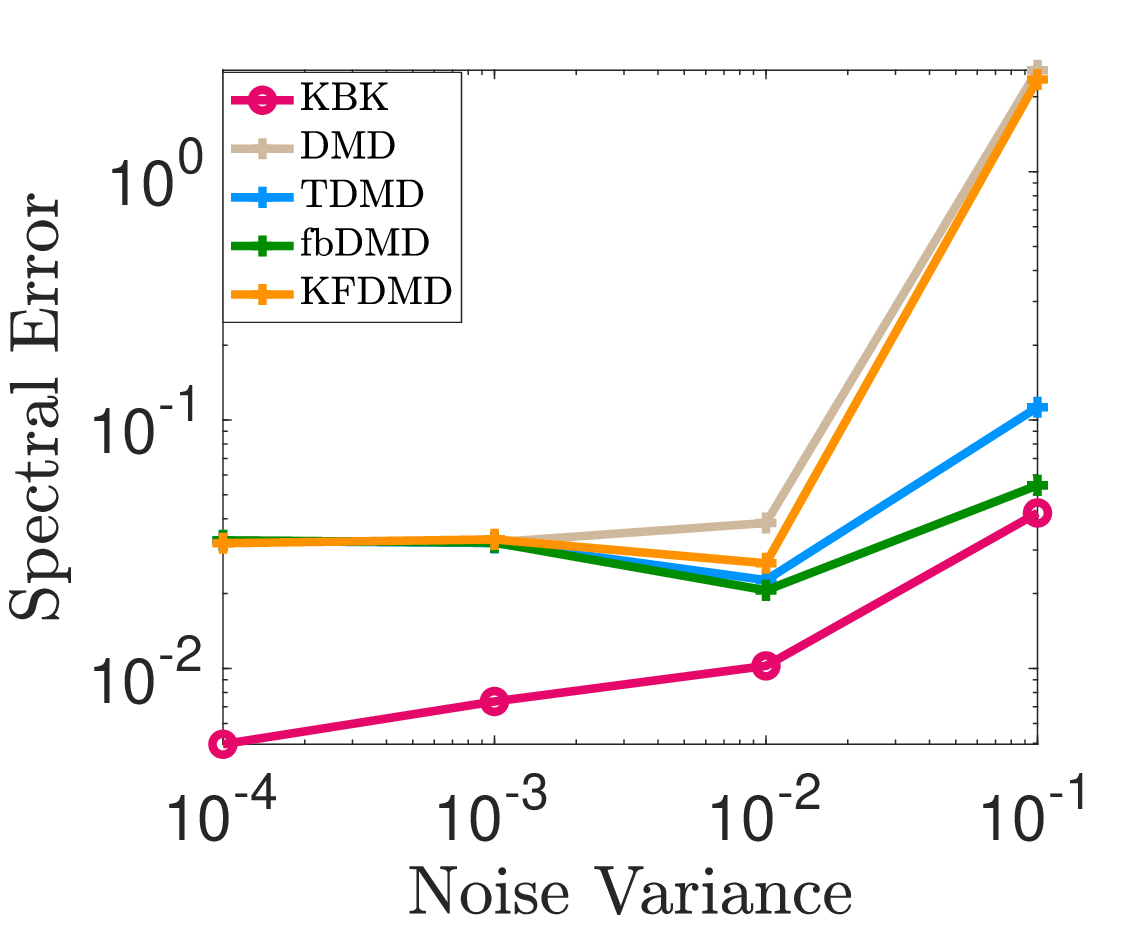}}
		\subfloat[]{
			\label{Fig1.sub3}
			\includegraphics[width=.3\textwidth]{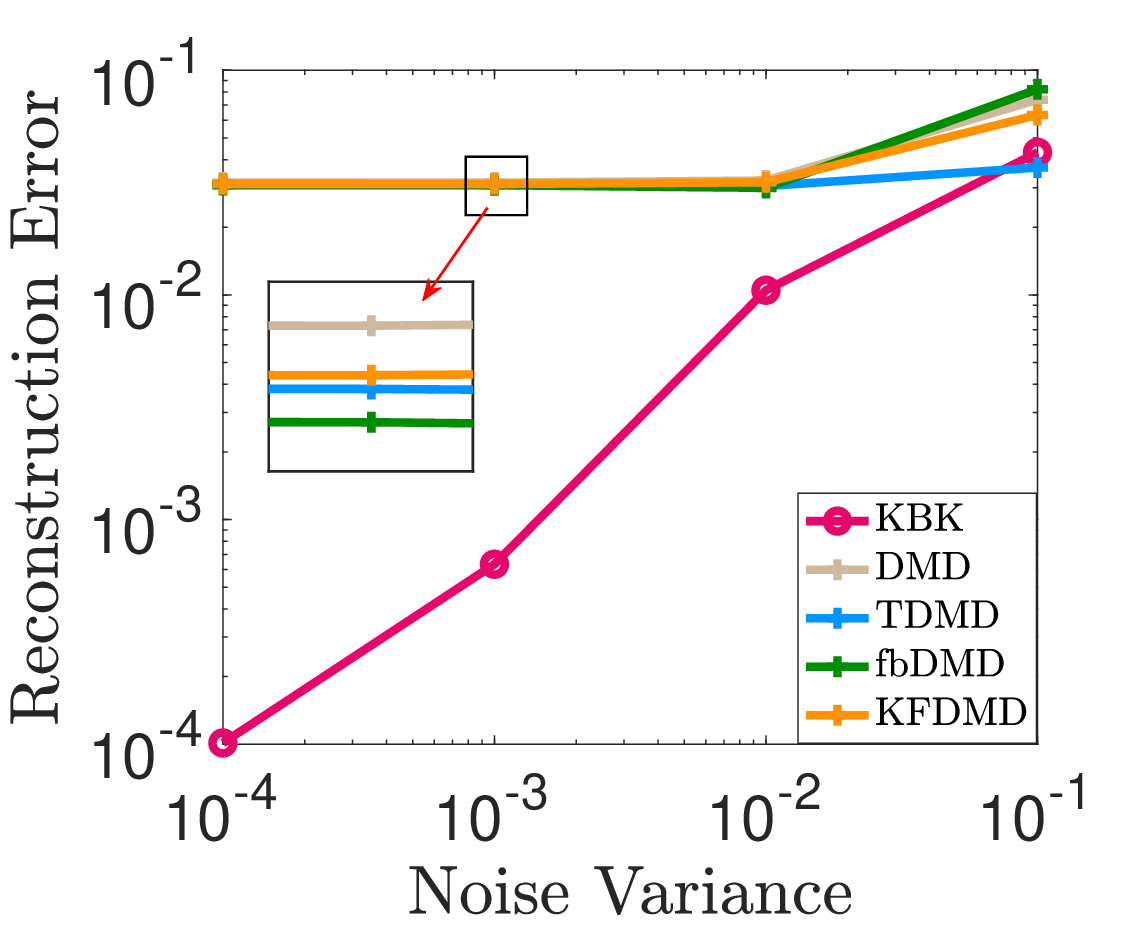}}

		\caption{The spectral analysis performance of the nonlinear system \eqref{eq:system1} using DMD, TDMD, fbDMD, KFDMD and KBK methods. (a) the spectral approximation when $\sigma^2 = 10^{-2}$, (b) the spectral approximation errors across varying noise variances $\sigma^2$, (c) the state reconstruction errors across varying noise variances $\sigma^2$.}
  
		\label{fig1}
	\end{figure*}

The principal Koopman eigenvalues of the system are $-1 + 3i$ and $-1 - 3i$. A total of $N=200$ data points were collected for analysis with $M=4$. The observations are contaminated by Gaussian noise $\mathcal{N}(0, \sigma^2)$, where $\sigma^2$ is set to $\{10^{-4},10^{-3},10^{-2},10^{-2}\}$. 
Fig. \ref{fig1} illustrates the experimental results for this dynamical system, comparing the performance of DMD, TDMD, fbDMD, KFDMD, and KBK methods under varying noise levels. In particular, Fig. \ref{Fig1.sub1} presents the eigenvalue approximations based on these methods at a noise level of $\sigma^2=10^{-2}$. It is evident that the DMD method performs poorly in noisy environments, resulting in inaccurate spectral estimates. While all methods yield approximate eigenvalues with imaginary parts closely matching the true values, for the real parts, the KBK method provides the most accurate spectral approximations. Its approximations align closely with the theoretical values, demonstrating superior performance in spectrum approximation under these noise conditions.

Fig. \ref{Fig1.sub2} presents the spectrum approximation error $E_1$ across various noise levels, where the noise variance $\sigma^2$ ranges from $10^{-4}$ to $10^{-1}$. The results indicate that, at lower noise levels, the performance of the other methods is largely comparable, with the KBK method demonstrating superior accuracy, as its approximation error is approximately one-tenth that of the other methods. Moreover, across all noise levels, the KBK method consistently demonstrates lower error compared to the other methods. This highlights the superior performance of the KBK method compared to alternative approaches for Koopman spectrum approximation.

Fig. \ref{Fig1.sub3} illustrates the state estimation error associated with various methods under different noise levels, with the error $E_2$ calculated using \eqref{eq:error_2}. The results indicate that when the noise variance is below $10^{-1}$, the KBK method significantly outperforms the other methods in terms of estimation accuracy. 
The results show that when the noise variance is below \(10^{-1}\), the KBK method significantly outperforms others in estimation accuracy, thanks to the Kalman smoothing that improves data filtering and precision. In contrast, the KFDMD method, despite using Kalman filtering, performs poorly due to model inaccuracies stemming from noisy data, and its effectiveness heavily relies on prior knowledge of noise characteristics. As noise levels rise, the KBK method's estimation error increases, and at a variance of \(10^{-1}\), its reconstruction error becomes comparable to TDMD, though it still outperforms the other methods.

The KBK method, which iteratively updates between Kalman smoothing and system identification, incurs higher computational costs than one-shot methods like DMD or TDMD. However, for the systems considered in this section, the overhead remains acceptable, with the entire procedure typically completing in about 5 seconds. This modest increase in computation time is well justified by the substantially improved accuracy of the estimated eigenvalues. For parameters, increasing \(N\) improves the accuracy but eventually reaches a plateau. Raising \( M \) generally reduces the MSE, though it may introduce minor oscillations in the error curve. Both adjustments increase computational cost and the number of iterations. A practical approach is to start with smaller values and gradually increment them until the additional gains become negligible. 

	\section{Conclusion}
	This study proposes a robust method for approximating the Koopman spectrum from noisy measurements. By integrating Bayesian learning with Kalman smoothing, it constructs a statistical framework that effectively mitigates the impact of measurement noise within the Koopman operator framework. The method enhances performance by iteratively updating both state estimation and operator identification. A notable strength of this approach lies in its robustness across varying noise level for Koopman spectrum approximation and states reconstruction, even though the noise level is unknown. This resilience under varying noise conditions is thoroughly validated through extensive comparative experiments.
 
 However, 
 for certain dynamical systems, the effectiveness of most Koopman spectral analysis methods is strongly dependent on the choice of the initial point. This sensitivity can lead to significant variations in spectral approximations. In future work, we aim to develop more robust methods that reduce this dependence, improving reliability for a wider range of dynamical systems.
	
	\bibliographystyle{IEEEtran}
	\bibliography{IEEEabrv,cite}

\end{document}